\documentclass[manuscript,screen,nonacm]{acmart}

\AtBeginDocument{%
  }

\setcopyright{acmcopyright}
\copyrightyear{2024}
\acmYear{2025}
\acmDOI{XXXXXXX.XXXXXXX}

\usepackage{caption}
\usepackage{subcaption}
\usepackage{multirow}
\usepackage{enumitem}
\usepackage{xspace}
\usepackage{tablefootnote}
\usepackage{float}
\usepackage{graphics}
\usepackage[normalem]{ulem}
\usepackage{xstring}
\usepackage{tabularx}
\newcolumntype{P}[1]{>{\centering\arraybackslash}p{#1}}

\DeclareCaptionFormat{cont}{#1 Continued#2#3\par}

\newcommand{\methodName}{SummAct\xspace}
\newcommand{\detailName}{UI element attention\xspace}

\newcommand{\datasetpc}{Mind2Web\xspace}
\newcommand{\datasetmobile}{MoTIF\xspace}

\begin{document}

\title[\methodName]{\methodName: Uncovering User Intentions Through Interactive Behaviour Summarisation}

\author{Guanhua Zhang}
\affiliation{%
  \institution{University of Stuttgart}
  \city{Stuttgart}
  \country{Germany}
}
\email{guanhua.zhang@vis.uni-stuttgart.de}

\author{Mohamed Ahmed}
\affiliation{
  \institution{University of Stuttgart}
  \city{Stuttgart}
  \country{Germany}
}
\email{st178773@stud.uni-stuttgart.de}

\author{Zhiming Hu}
\affiliation{%
  \institution{University of Stuttgart}
  \city{Stuttgart}
  \country{Germany}
}
\email{zhiming.hu@vis.uni-stuttgart.de}

\author{Andreas Bulling}
\affiliation{%
  \institution{University of Stuttgart}
  \city{Stuttgart}
  \country{Germany}
}
\email{andreas.bulling@vis.uni-stuttgart.de}

\renewcommand{\shortauthors}{Zhang et al.}

\begin{abstract}
Recent work has highlighted the potential of modelling interactive behaviour analogously to natural language.
We propose \textit{interactive behaviour summarisation} as a novel computational task and demonstrate its usefulness for automatically uncovering latent user intentions while interacting with graphical user interfaces.
To tackle this task, we introduce \methodName~-- a novel hierarchical method to summarise low-level input actions into high-level intentions.
\methodName first identifies sub-goals from user actions using a large language model and in-context learning.
High-level intentions are then obtained by fine-tuning the model using a novel
\detailName to preserve detailed context information embedded within UI elements during summarisation.
Through a series of evaluations, we demonstrate that \methodName significantly outperforms baselines across desktop and mobile interfaces as well as interactive tasks by up to 21.9\%.
We further show three exciting 
interactive applications benefited from \methodName: interactive behaviour forecasting, automatic behaviour synonym identification, %
and language-based behaviour retrieval.

\end{abstract}


\maketitle

\section{Introduction}

Recent work has demonstrated that interactive behaviour, e.g. when interacting with graphical user interfaces using the mouse or keyboard, shares similarities with the sequential and hierarchical nature of natural language~\cite{zhang2023exploring}.
In parallel, an increasing number of works have started to model interactive behaviour as natural language %
and process it using language models~\cite{huang2024automatic,li2024omniactions, wen2023empowering, deng2023mind2web, li2020mapping}.
One key advantage of this language perspective is facilitating a more interpretable analysis and understanding of interactive behaviour, thus enabling novel paradigms for solving human-computer interaction (HCI) tasks.

Among these tasks, understanding users' intentions is key to intelligent interactive systems and anticipatory user interfaces~\cite{zhang2023exploring, hu2022EHTask}.
Recognising intentions based on the user's behaviour history has been widely studied and applied in HCI, including for unintentional error detection~\cite{almehmadi2021micro}, next action prediction~\cite{bednarik2012you}, or task automation~\cite{zheng2024gpt, wen2023empowering, humphreys2022data}.
Despite its potential for HCI and promising first results, predicting users' intentions from their interactive behaviour remains challenging, partly due to human behaviour's high variability and complexity.
Existing works, therefore, typically assume a pre-defined and fixed set of intentions and treat intention recognition as a classification task.
However, this approach neither captures the wide variety of user intentions in everyday scenarios
nor can robustly adapt to unseen or context-dependent intentions~\cite{yuan2024generating}.
It can also result in misinterpretations when users' needs do not align with predefined intention categories, which often happens in real-world applications~\cite{zhang2022predicting}.

In this work, we take inspiration from text and image summarisation tasks studied in natural language processing and computer vision.
These tasks involve summarising long text or complex videos into a concise sentence description.
Similarly, we formulate intention recognition as an interactive behaviour summarisation task: human interactive behaviour is to be summarised into a sentence, i.e., a natural language description of users' underlying interactive intentions.
In contrast to existing methods, interactive behaviour summarisation enables recognising an open-ended set of intentions. It allows for capturing more flexible and varied interaction intentions and handling unseen intentions.

To address this task, we propose \methodName ~-- a novel large language model (LLM)-based method that uses a hierarchical summarisation process: the method initially summarises \textit{low-level} actions into \textit{mid-level} sub-goals, then uses them to augment the input, and finally summarise into a \textit{high-level} intention.
In this work, we focus on the interactive behaviour at the user interface (UI) element level, meaning that each input action sample consists of the interacted UI element and the operation the user conducts on this element (e.g., click or select).
The UI element information includes its category (e.g., button or combo box), inherent (e.g., the name or the visible text on a button) and additional content (further values that users are interested and pick, e.g., a value selected from a combo box).
On these input actions, \methodName first generates sub-goals using in-context learning via a pretrained, frozen LLM due to the lack of ground-truth annotations and then
fine-tunes the LLM to produce the final summary.
During fine-tuning, we further propose a \detailName mechanism that assigns higher weights to the %
UI element contents, thereby preserving the detailed context information embedded within these elements. %
This is crucial for accurately interpreting intentions that exhibit subtle differences and for further applications like behaviour forecasting.

We evaluate \methodName on two datasets that cover a web (\datasetpc~\cite{deng2023mind2web}) and a mobile (\datasetmobile~\cite{burns2022motifvln}) interaction setting.
We show that \methodName can accurately uncover the intentions underlying user actions, with a sentence embedding cosine similarity up to 0.842 compared to the ground-truth intentions.
We also demonstrate the importance of our design choices with the full \methodName model significantly outperforming ablated versions by up to 21.9\% in cosine similarity.
We finally showcase three exciting applications enabled by interactive behaviour summarisation: providing contextual information of user intentions to enhance behaviour forecasting for proactive user interfaces;
automatically identifying behaviour synonyms to understand user preferences, interaction strategies, system usability and common design patterns;
and unlocking language-based behaviour retrieval that lays the foundation for building behaviour-related conversational agents. %

\noindent
In summary, the specific contributions of our work are three-fold:
\begin{itemize}[leftmargin=15pt]
    \item We formulate intention recognition %
    as the novel open-ended task of summarising interactive behaviour into natural language descriptions.
    This formulation overcomes existing limitations associated with pre-defined intention sets and improves generalisability to unseen intentions.
    Towards this task, we propose an LLM-based method, \methodName\footnote{We will release our source code upon acceptance.} incorporating two distinct and novel designs~--~ hierarchical summarisation and \detailName. %

    \item We show the effectiveness of these designs, and \methodName in general, for interactive behaviour summarisation, through a series of evaluations on two datasets covering both web and mobile interaction settings.
    
    \item We demonstrate the potential of interactive behaviour summarisation via three example applications: interactive behaviour forecasting, identifying behaviour synonyms, and language-based behaviour retrieval. These are widely relevant in HCI, particularly for developing intelligent interactive systems or UI optimisation.
    
\end{itemize}

\section{Related work}
We discuss related work on (1) understanding user intentions behind interactive behaviour, (2) large language models for interactive behaviour modelling, and (3) summarising non-language data.

\subsection{Understanding User Intentions behind Interactive Behaviour}
Recognising users' intentions from their interactive behaviour is key to intelligent user interfaces, which can proactively support users by automatically adjusting the UIs or providing action recommendations~\cite{fu2017your,zhang2022predicting}.
Therefore, an increasing number of works in HCI field have studied automatic intention prediction from interactive behaviour.
For example, in virtual reality (VR) interactive environments, David-John et al.~\cite{david2021towards} recognised user intentions of selecting an item from gaze behaviour.
Hu et al.~\cite{hu2022EHTask} built a model based on convolutional neural networks and bidirectional gated recurrent units to recognise interactive tasks the users aimed to achieve from their eye and head movements in VR.
In more pervasive, daily scenarios such as interacting with personal computers or mobile devices, researchers have also developed various methods to recognise the objectives users aim to achieve via their actions.
These actions were captured through different modalities, including mouse movements and clicks, keyboard typing, eye tracking and touch interactions.
For instance,
Elbahi et al.~\cite{elbahi2013hidden} used hidden Markov model and conditional random field to recognise which e-learning task the users were performing from mouse movement.
Koldijk et al.~\cite{koldijk2012real} developed classifiers based on different machine learning models including naive Bayes, KStar, decision tree and multilayer perceptron (MLP) to recognise which one out of 12 office tasks the user aimed to complete from their mouse and keyboard actions.
Zhang et al.~\cite{zhang2022predicting} proposed a multimodal random forest-based approach to recognise intentions from users' mouse, keyboard and gaze actions in a text editing task that included seven text formatting intentions pre-defined by the authors.
In the mobile interaction settings, Xu et al.~\cite{xu2020recognizing} identified intentional versus unintentional touches from gaze, head and screen touch behaviour.
They built the model based on logistic regression, naïve Bayes, k-nearest neighbour, random forest, gradient boosting and MLP.

However, all the above works studied a pre-defined, fixed, and closed set of user intentions, which inherently limits the adaptability and scalability of systems to new intentions.
Inspired by the prior finding that interactive behaviour shares a similar sequential and hierarchical structure with natural language~\cite{zhang2023exploring}, 
in this paper, we formulate intention recognition as summarising interactive behaviour into a sentence.
This attempt accommodates an open-ended set of user intentions and thus allows for more flexible and comprehensive interactive behaviour modelling.

\subsection{Large Language Models for Interactive Behaviour Modelling}
LLMs have recently achieved ground-breaking success in HCI research, bringing novel insights and methodology to model user actions for different applications.
For example, 
Liu et al.~\cite{liu2024unblind} proposed HintDroid, an LLM-based method using in-context learning to generate hint-text in Android applications based on the user's input and its corresponding UI context.
Wang et al.~\cite{wang2023enabling} used a pretrained LLM to investigate the conversational interactions with mobile user interfaces via prompt engineering and zero-shot learning.
Their results demonstrated the potential of using LLMs for language-based mobile interactions.
Huang et al.~\cite{huang2024automatic} applied pretrained LLMs and a chain-of-thought technique to extracting macros from mobile interaction traces in existing datasets.
Other research focuses on building LLM-based automatic agents that navigate through interactive systems and complete pre-defined tasks~\cite{li2020mapping,cheng2024seeclick,zheng2023synapse,wen2023empowering}.
For instance, Deng et al. proposed MindAct~\cite{deng2023mind2web} to automatically perform given tasks in complex web environments.
MindAct first fine-tuned a language model to rank all the UI elements available on the web page based on the task description and action history.
Selecting the top-ranked as the candidates, MindAct then formulated task automation as a multi-choice question-answering task and used in-context learning for task automation.

Despite the acknowledged potential of LLMs in interactive behaviour modelling, their application in intention recognition remains largely under-explored.
In this paper, we approach intention recognition through an interactive behaviour summarisation task and propose an LLM-based method, \methodName, to address this task.

\subsection{Summarising Non-Language Data}
In language processing, summarisation has been widely studied and applied in condensing large amounts of information into concise sentences, enabling efficient content consumption across various domains, such as documents~\cite{liu2019hierarchical,el2021automatic}, code~\cite{haque2022semantic}, and speech~\cite{rezazadegan2020automatic}.
Inspired by the success of these works, researchers have started to summarise non-language data also into coherent textual descriptions for quick and accessible understanding of complex data~\cite{wang2023enabling}.
For instance, image captioning enhanced the comprehension of the images and meanwhile enables text-based image retrieval~\cite{rotstein2024fusecap}.
Kawamura et al.~\cite{kawamura2024fastperson} proposed a multimodal method to summarise lecture videos using audio transcripts, on-screen images and texts enabling users to effectively obtain information from lengthy video content.
Lin et al.~\cite{lin2024motion} and Chen et al.~\cite{motionllm} summarised human motion videos, which not only enhanced the understanding of the motion sequence, but also had the potential to allow controllable text-to-motion generation.
Chen et al.~\cite{chen2024gazexplain} annotated natural-language explanations for fixations in scanpaths, providing insights into implicit gaze behaviour change and benefited explainable scanpath prediction.
In HCI, researchers have studied the summarisation of graphical user interfaces.
Wang et al.~\cite{wang2021screen2words} summarised core information of mobile UI screens into natural language via the proposed multimodal method, Scree2Words, integrating the text, image, structures and UI semantics.
They showcased that the summarisation could potentially be used for language-based UI retrieval, enhancing screen readers and screen indexing for conversational mobile interactions.

These applications highlight the transformative potential of applying summarisation techniques to diverse data modalities.
Building upon this, our work introduces a novel method to summarise the complex interactive behaviour into human-interpretable natural language sentences, which reflect users' latent intentions.
Additionally, we present three examples benefited from interactive behaviour summarisation, behaviour forecasting, behaviour synonym identification and language-based behaviour retrieval, which are widely relevant for intelligent interactive systems and user interface optimisation.

\section{Interactive behaviour summarisation using \methodName}
Building on recent advances demonstrating the potential of analysing interactive behaviour similarly to natural language~\cite{zhang2023exploring, li2020mapping, huang2024automatic}%
, our method \methodName addresses the novel task of
interactive behaviour summarisation.
The input of \methodName is a sequence of UI element-level interaction actions caused by the user interacting with a graphical user interface.
Every action consists of the UI element the user has interacted with and the operation performed on this element (e.g., click or select).
The UI element contains the information of its category (e.g., button or combo box), the inherent (e.g., its name or the text on it) and additional content (the specific value the user is interested in, e.g., the value selected from a combo box).
The output of our method is a natural language sentence that concisely summarises their interactive behaviour and, as we show here, their latent interaction intentions.
In stark contrast to existing methods for the classification of intentions, which are limited to a closed and predefined set of possible intentions~\cite{elbahi2013hidden, elbahi2015web, zhang2024mouse2vec}, interactive behaviour summarisation allows us to recognise an open-ended set of intentions, including intentions not seen during training. %
As such, \methodName can provide a more comprehensive and generalisable understanding of interactive behaviour.

\begin{figure*}
    \centering
    \includegraphics[width=.9\textwidth]{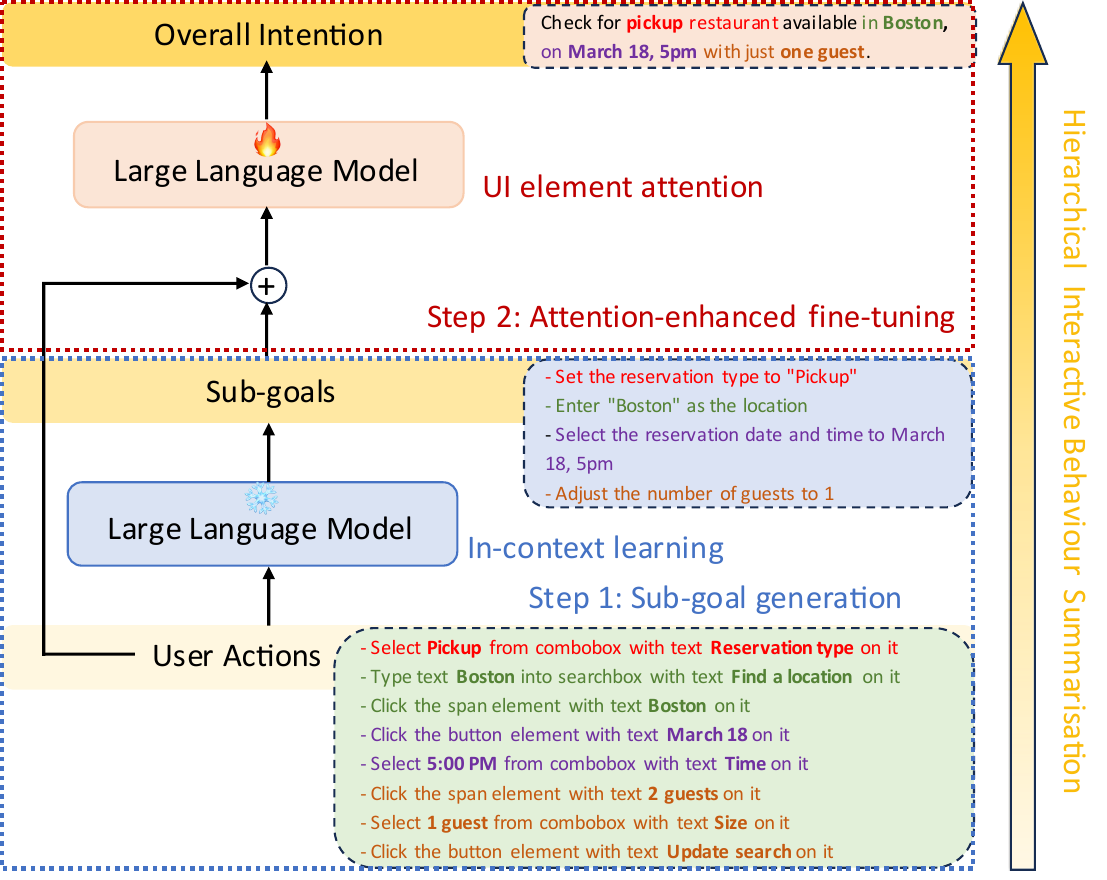}
    \caption{Overview of \methodName for uncovering user intentions during user interface interactions through interactive behaviour summarisation.
    \methodName employs a hierarchical process that initially generates sub-goals and produces the overall intention in natural language.
    The input is a sequence of user actions, including the interacted UI element and the user's operation on this element.
    \methodName uses in-context learning to infer an arbitrary number of sub-goals using a pretrained, frozen LLM (Step 1) and then fine-tunes the LLM while introducing a \detailName (Step 2) to keep detailed context embedded in UI element contents, as highlighted in \textbf{bold}.
    Actions in the same colour are summarised into the same sub-goal and then to a phrase in the overall intention.
    The summary of the output reflects the latent intentions that underlie these actions.
    }
    \Description{X}
    \label{fig:overview}
\end{figure*}

\autoref{fig:overview} provides an overview of \methodName's hierarchical approach to interactive behaviour summarisation:
given a sequence of user input actions encoded in natural language descriptions, \methodName first summarises these low-level actions into a set of sub-goals.
Due to the absence of ground-truth data, we use expert annotations in combination with in-context learning to adapt a pretrained, frozen LLM to generate sub-goals.
In the second step, these sub-goals are combined with the original actions and summarised into high-level intentions via fine-tuning the LLM.
To
preserve UI element content 
(indicated in \textbf{bold} in~\autoref{fig:overview}) in the summary, our method uses a novel \detailName during fine-tuning.
Two previous findings inspire this hierarchical approach: hierarchical modelling of language data can robustly handle extensive and complex input, such as long documents~\cite{liu2019hierarchical};
and interactive behaviour has an inherent hierarchical nature similar to that observed in natural language~\cite{zhang2023exploring}.
In the following, we describe each of these steps in more detail.

\subsection{Sub-Goal Generation}
\label{sec:method-subintention}

The first step involves generating sub-goals from the low-level, individual input actions.
As shown in ~\autoref{fig:overview}, actions marked in the same colour are summarised into the same sub-goal, which later becomes a phrase integrated into the overall intention.
Given the lack of HCI datasets offering annotations of interaction sub-goals, we used in-context learning.
In-context learning involves giving an LLM a small set of examples presented within the context (the prompt) at inference time to guide its response~\cite{wei2019eda}.
This approach leverages LLM's ability to understand and adapt to patterns presented in the immediate context of the query without the need to fine-tune the model.
To obtain these examples, we asked three HCI, GUI, and behaviour modelling experts to annotate the sub-goals on five samples from the training set~\cite{lihuman}.
These samples are five different action sequences completing five different tasks.
In Appendix ~\ref{sec:prompt-subintention}, we provide the used prompt, including the example of sub-goal annotation.%

\subsection{Attention-Enhanced Fine-Tuning}
\label{sec:method-loss}

In the second step, we fine-tune the LLM to summarise the overall intention from the generated mid-level sub-goals and the original low-level actions.
In Appendix ~\ref{sec:prompt-finetuning}, we provide sample prompts used in this fine-tuning step.
LLMs are typically structured as sequence-to-sequence models, i.e. they are trained to generate output sequences based on input sequences, such as summarising an input.
Therefore, LLMs are commonly trained using a next token prediction task in a teacher-forcing setup, where the model is guided by a ground-truth token rather than the previously predicted token to predict the next token~\cite{qian2024tell}.
This training strategy helps stabilise the training process and accelerates convergence by reducing the propagation of errors through the sequence~\cite{wang2023enabling}.
Thus, based on the input prompt, LLMs iteratively predict the next token and continually update their predictions as each new token is added to the output sequence.
Next token prediction is formulated as a classification task 
and thus uses a cross-entropy loss $L_{NextToken}$~\cite{li2021mixed}.
For the $j$-th token in the input sequence, this loss is calculated as
\begin{equation}
    L_{NextToken_j} = log(\mathbb{P}_\theta(Token_j | Token_1, ..., Token_{j-1}))
\end{equation}
where $\theta$ are the model parameters.

In preliminary experiments, we found that fine-tuning the LLM only using next token prediction led to %
UI element content is getting excluded from the final interactive behaviour summary.
This may be because the model tends to rely on frequent patterns of general natural language rather than focus on task-specific information embedded in the UI elements~\cite{bachmann2024pitfalls}.
\autoref{fig:overview} shows examples of such information related to the interactive behaviour summarisation task (highlighted in \textbf{bold}).
Let us consider the first input action ("Select \textbf{Pickup} from combobox with text \textbf{Reservation type} on it") as an example: "Reservation type" is the name of the combo box, i.e., the inherent content of the combo box, representing what this combo box is about; while "Pickup" is an additional content of the combo box, namely one value the combo box provides that the user is interested in and ultimately selects.
Retaining such UI element contents in the final summaries is particularly important for interactive behaviour summarisation:
First, such content provides interactive context information necessary to distinguish between subtle intentions.
For instance, actions include selecting ``1'' from a combo box named ``guest number'' on a booking site by some users versus selecting more guests by other users.
However, these detailed contents are ignored, and the summarised intentions are both \textit{finding a hotel room}; the system may further inaccurately suggest unsuitable accommodations, e.g., family rooms for solo travellers and vice versa, causing decreased usability and potential frustration.
Second, they are important for downstream applications, such as behaviour forecasting, to ensure the prediction is relevant and consistent with the current context and underlying goals~\cite{wen2023empowering}.
For example, if a user clicks on a button named "gluten-free" but this content is overlooked, the interactive system may mistakenly predict the upcoming actions to involve browsing or purchasing products containing gluten, leading to a worse user experience.

To address this challenge
we propose a \detailName mechanism to enhance the fine-tuning process by guiding the model to focus on these contents.
This is similar to ensuring that a text summary covers essential keywords in natural language processing~\cite{el2021automatic}.
Specifically, for the $i$-th training sample, we create an attention vector $\mathbf{K}_i$, where each component $K_{ij}$ denotes the amount %
of attention assigned to the $j$-th token in the ground-truth summary:
\begin{equation}
K_{ij} = 
\begin{cases} 
\lambda & \text{if \,\,\,} Token_j \in Token_{Detail} \\
1 & \text{otherwise}
\end{cases}
\end{equation}
As such, tokens that contain action details receive $\lambda$ times the attention compared to other tokens.
We empirically set $\lambda = 2$
in our experiments.
The overall fine-tuning loss integrating this attention mechanism $\mathcal{L}_{Enhanced}$ is then computed as a weighted version of the original cross-entropy loss:
\begin{equation}
    \mathcal{L}_{Enhanced} = \mathbf{K} \circ \mathcal{L}_{NextToken}
\end{equation}

\subsection{Implementation}
\label{sec:method-implementation}

We opted for the lightweight open-source Mistral-7B model~\cite{jiang2023mistral} as the LLM backbone, known for its efficiency and effectiveness in handling various NLP tasks.
Mistral-7B incorporates advanced techniques such as grouped-query attention for fast inference and sliding window attention for managing long sequences and complex contexts.
These features contribute to Mistral-7B's superior performance on various benchmarks compared to other state-of-the-art models while using fewer parameters, thus conserving computational resources~\cite{pentyala2024paft}.
We used a batch size of 16 and a maximum input sequence length of 1024, with an initial learning rate of 1e-6. %
We used the Adam optimiser with $\beta_1=0.9$ and $\beta_2=0.95$~\cite{qian2024tell}, and a cosine annealing scheduler for a progressive reduction of the learning rate following a cosine curve, a strategy proven to stabilise the training phase~\cite{loshchilov2016sgdr}.
We fine-tuned the model for 15 epochs using eight Tesla V100-SXM2-32GB GPUs, completing the training within ten hours.

\section{Experiments}

We conducted experiments to evaluate the quality of interactive behaviour summaries generated by \methodName.
Given the novelty of this task and the lack of existing baseline methods, we compare the full model with several ablated versions instead.
More specifically, starting with using an off-the-shelf, pretrained LLM -- the common practice in HCI research currently~\cite{huang2024automatic, berkovitch2024identifying} -- we incrementally add fine-tuning, sub-goal generation, and the \detailName mechanism.
We report quantitative metrics that measure how similar the generated summaries are compared to the ground truth, as well as qualitative similarities and differences of the generated summaries.

\subsection{Datasets}
\label{sec:eval-datasets}

We conducted all evaluations using two prominent datasets that encompass both the web (\datasetpc~\cite{deng2023mind2web}) and mobile (\datasetmobile~\cite{burns2022motifvln}) interaction contexts.
These datasets are extensively used to understand and model user interfaces and interactive behaviours~\cite{berkovitch2024identifying,zheng2023synapse}.
They encode interactive behaviour as user actions to achieve specified interaction objectives.
Each user action is annotated with information related to the UI element (category and content) and the user operation (e.g., click or swipe) associated with this element.

\subsubsection{\datasetpc}
This dataset provides crowdsourced actions across 2,350 tasks performed on 137 real-world websites (e.g., Booking, Uniqlo, IMDB) spanning 31 domains (e.g., travel, shopping, entertainment).
As such, this dataset offers a wide variety of user actions and intentions and allows us to evaluate the performance of \methodName in real-world scenarios.
\datasetpc includes three different data subsets:
1) \textit{cross-domain} includes data instances from different domains, e.g., shopping vs travel; 2) \textit{cross-website} includes instances from unseen websites, e.g., Booking vs Airbnb; and 3) \textit{cross-task} includes unseen tasks, e.g., booking a flight vs buying a shirt.
We preprocessed the dataset by generating natural language descriptions for the provided raw actions using a transformation template~\cite{niu2024screenagent} (see ~\autoref{sec:appendix-description} for more details).

\subsubsection{\datasetmobile}

This dataset targets a
mobile interaction setting comprising screen touch data collected on 756 different tasks across 125 Android applications.
The dataset directly provides synthetic natural language sentences describing each low-level action
including the interacted UI element and the user's operation (click, type or swipe) on it.
\datasetmobile includes both feasible and infeasible tasks such as tasks that are too unclear or cannot be completed in the given App. %
We only used the feasible tasks for our evaluations to ensure that user actions reliably reflect the corresponding intentions.

\subsection{Ablations}
We compared the full \methodName model \textbf{LLM+FT+SubGoal+Attn} (fine-tuned LLM using both the input actions and sub-goals with $\mathcal{L}_{Enhanced}$ for summarisation) with several ablated versions to evaluate the impact of the different modifications.
To ensure a fair comparison, all methods used the same LLM and the same prompt templates.
\begin{itemize}[leftmargin=15pt]
    \item \textbf{LLM}: pretrained LLM using only the input actions.
    \item \textbf{LLM+SubGoal}: pretrained LLM using both the input actions and the sub-goals.
    \item \textbf{LLM+FT}: fine-tuned LLM using only the input actions with $\mathcal{L}_{NextToken}$ for summarisation.
    \item \textbf{LLM+FT+Attn}: fine-tuned LLM using only the input actions with $\mathcal{L}_{Enhanced}$ for summarisation.
    \item \textbf{LLM+FT+SubGoal}: fine-tuned LLM using both input actions and sub-goals with $\mathcal{L}_{NextToken}$ for summarisation.
\end{itemize}
Since the \detailName mechanism is specifically incorporated into the loss function of fine-tuning, we do not have a standalone version of the pretrained LLM enhanced solely by the attention, i.e., LLM+Attn.

\subsection{Quantitative Evaluations}
\label{sec:eval-quantitative}

\begin{table*}
\begin{tabularx}{\textwidth}{l X c c c P{2cm}}
\toprule
\multirow{2}{*}{\textbf{Method}} & \multirow{2}{*}{\textbf{Metric}} & \multicolumn{3}{c}{\textbf{\datasetpc}} & \multicolumn{1}{c}{\multirow{2}{*}{\textbf{\datasetmobile}}} \\\cline{3-5}
 &  & \textbf{Cross Domain} & \textbf{Cross Task} & \textbf{Cross Website} & \multicolumn{1}{l}{} \\\midrule
\multirow{4}{*}{\textbf{LLM}} & CosSim &.357 &.348 &.380 &.203 \\
 & BLEU &.004 &.004 &.004 &.012 \\
 & ROUGE &.050 &.056 &.060 &.077 \\
 & METEOR &.126 &.132 &.146 &.093 \\\midrule
\multirow{4}{*}{\textbf{LLM+SubGoal}} & CosSim &.381 &.374 &.404 &.230 \\
 & BLEU &.006 &.004 &.006 &.009 \\
 & ROUGE &.051 &.055 &.059 &.063 \\
 & METEOR &.150 &.157 &.167 &.098 \\\midrule\midrule
\multirow{4}{*}{\textbf{LLM+FT}} & CosSim &.631 &.661 &.673 &.691 \\
 & BLEU &.201 &.217 &.204 &.293 \\
 & ROUGE &.313 &.301 &.312 &.511 \\
 & METEOR &.303 &.306 &.322 &.510 \\\midrule
\multirow{4}{*}{\textbf{LLM+FT+Attn}} & CosSim &.705 &.740 &.756 &.785 \\
 & BLEU &.349 &.296 &.305 &.407 \\
 & ROUGE &.345 &.372 &.383 &.730 \\
 & METEOR &.293 &.341 &.374 &.709 \\\midrule
\multirow{4}{*}{\textbf{LLM+FT+SubGoal}} & CosSim &.718 &.753 &.763 &.756 \\
 & BLEU &.291 &.301 &.301 &.359 \\
 & ROUGE &.381 &.392 &.391 &.689 \\
 & METEOR &.323 &.332 &.363 &.662 \\\midrule\midrule
\multirow{4}{*}{\textbf{\begin{tabular}[c]{@{}l@{}}LLM+FT+SubGoal+Attn \\ (\methodName)\end{tabular}}} & CosSim & \textbf{.755} & \textbf{.796} & \textbf{.802} & \textbf{.842} \\
 & BLEU & \textbf{.406} & \textbf{.453} & \textbf{.445} & \textbf{.453} \\
 & ROUGE & \textbf{.390} & \textbf{.432} & \textbf{.447} & \textbf{.799} \\
 & METEOR & \textbf{.376} & \textbf{.430} & \textbf{.454} & \textbf{.758}\\
 \bottomrule
\end{tabularx}
FT = Fine-tuning; SubGoal = Sub-goal generation; Attn = \detailName
\vspace{.3cm}
\caption{Interactive behaviour summarisation results achieved by our proposed \methodName and its ablated versions.
The evaluation is conducted on a web dataset \datasetpc (including three test subsets for generalisability assessment across domains, tasks and websites) and a mobile dataset \datasetmobile.
We measure the summarisation quality with four metrics, cosine similarity between sentence embeddings, and n-gram based BLEU, ROUGE and METEOR.
The best results are shown in \textbf{bold}.
}
\label{tab:eval-results}
\end{table*}

We first quantify the similarity between ground truth and summarised intentions for all methods with four widely used NLP metrics~\cite{wang2023enabling,zhang2024systematic}.
Specifically, we report Recall-Oriented Understudy for Gisting Evaluation (ROUGE)~\cite{lin2004rouge},
Metric for Evaluation of Translation with Explicit ORdering (METEOR)~\cite{banerjee2005meteor}, and
Bilingual Evaluation Understudy (BLEU) ~\cite{papineni2001bleu} %
that count the overlapping n-grams between texts to assess their similarity, thus providing a measure that reflects lexical precision and recall.
We further report an embedding-based metric using
a state-of-the-art sentence encoder, Sentence Transformer\footnote{\url{https://huggingface.co/sentence-transformers/all-MiniLM-L6-v2}}, to obtain the sentence embeddings.
Embedding-based metrics evaluate the cosine similarity between sentences
and capture deeper and more robust semantic meanings that go beyond mere lexical matches~\cite{7995948}.
All of these metrics indicate better results as their values increase.

\autoref{tab:eval-results} provides an overview of the results of this comparison.
As can be seen from the table, our proposed full model consistently outperforms the ablated versions on all test sets and all metrics, obtaining a cosine similarity of up to 0.842 with the ground-truth user intentions.
Among the three generalisation test sets from \datasetpc, \methodName performed better in \textit{cross-website} and worse in the \textit{cross-domain} setting.
The former is likely because different websites within the same domain share similar UI designs and thus require similar navigation patterns~\cite{deng2023mind2web}, which can be efficiently captured and integrated by our \methodName.
These results also show, however, that generalisation across domains remains challenging due to the variations in context and user interactions.

We can also see from \autoref{tab:eval-results} that directly using a pretrained LLM performs the worst while adding fine-tuning, sub-goals, or the \detailName mechanism improved performance notably.
Although adding sub-goals (LLM+SubGoal) increased cosine similarity by up to 13.3\% (0.203 vs 0.230 on \datasetmobile), the largest performance increase was achieved when adding fine-tuning (LLM vs LLM+FT), where the cosine similarity improved by up to 89.9\% (0.348 vs 0.661, cross-task setting) on \datasetpc, and 240.4\% (0.203 vs 0.691) on \datasetmobile.

Also, adding our two novel designs of sub-goals and \detailName contributes to the effectiveness of \methodName (LLM+FT vs LLM+FT+SubGoal+Attn), together leading to an up to 21.9\% improvement on the cosine similarity (0.691 vs 0.842 on \datasetmobile).
Comparing our method with LLM+FT+Attn, the \detailName mechanism increased the cosine similarity on \datasetpc by 7.1\% (0.705 vs 0.755) in the cross-domain setting, 7.6\% (0.740 vs 0.796) in the cross-task setting, 6.1\% (0.756 vs 0.802) in the cross-website setting, and by 7.3\% (0.785 vs 0.842) on \datasetmobile.
On n-gram-based metrics, \methodName obtained improvements of up to 53.0\% (0.296 vs 0.453 on \datasetpc cross-task setting) on BLEU, 16.7\% (0.383 vs 0.447 on \datasetpc cross-website setting) on ROUGE, and 28.3\% (0.293 vs 0.376 on \datasetpc cross-domain setting) on METEOR.
In Section~\ref{sec:app-NA}, we further show that
a lack of the UI contents harms performance for behaviour forecasting.
Similarly, \methodName outperformed its ablated version that removed the sub-goals (LLM+FT+Attn), where the cosine similarity increased by 5.2\% (0.718 vs 0.755) in \datasetpc cross-domain setting, 5.7\% (0.753 vs 0.796) in the cross-task setting, 5.1\% (0.763 vs 0.802) in the cross-website setting and 11.4\% (0.756 vs 0.842) on \datasetmobile.
Moreover, BLEU improved by up to 50.5\% (0.301 vs. 0.453), achieved on \datasetpc cross-task setting; the maximum enhancement of ROUGE 15.97\% (0.689 vs. 0.799) on \datasetmobile; while the largest improvement of METEOR reached 29.5\% (0.332 vs. 0.430), obtained in \datasetpc cross-task setting.
Taken together, these evaluations show the effectiveness of the proposed components
for interactive behaviour summarisation. 

\subsection{Qualitative Analysis}
\label{sec:eval-qualitative}

We further examined the summaries generated by \methodName and its ablations qualitatively to understand the impact of the different designs.

\subsubsection{Impact of \detailName}
Compared to the summaries generated by the full \methodName implementation, the ablated version without \detailName lacks detailed context information embedded in UI element contents.
For example, the ground-truth intention to \textit{Find a \textbf{campground} in Orlando for two adults to check in on \textbf{Mar 29} and check out on \textbf{Mar 30}} was correctly summarised by \methodName as \textit{Find a \textbf{campground} in Orlando for two adults from \textbf{March 29} to \textbf{March 30}}.
On the contrary, the ablated version (LLM+FT+SubGoal) produced a less accurate summary, \textit{Find \textbf{hotels} in Orlando for two adults in \textbf{March}}, missing the information of the precise dates and the specific type of accommodation.
This issue arose because during the summarisation, the ablation ignored the detailed content in a clicking action on the button of ``Find a KOA'', which specified the accommodation type as a campground instead of any hotel.

Another example is the intention \textit{Find a highest rated dealer for Cadillac with a rating above \textbf{4 stars} within \textbf{20 miles} of \textbf{zip 60606}}.
\methodName effectively summarised this into \textit{Find a highest rated Cadillac dealer above \textbf{4 star} within \textbf{20 miles} of \textbf{60606}}.
However, the ablation's prediction was simply \textit{Find the highest rated dealer for Cadillacs}, missing the specific criteria of rating and proximity.

This analysis shows that without the \detailName mechanism, although the summaries retain the overall logic, they lack crucial specific information provided by the UI elements.

\subsubsection{Impact of sub-goals}

\begin{figure*}
    \centering
    \begin{subfigure}[t]{\linewidth}
        \includegraphics[width=\linewidth]{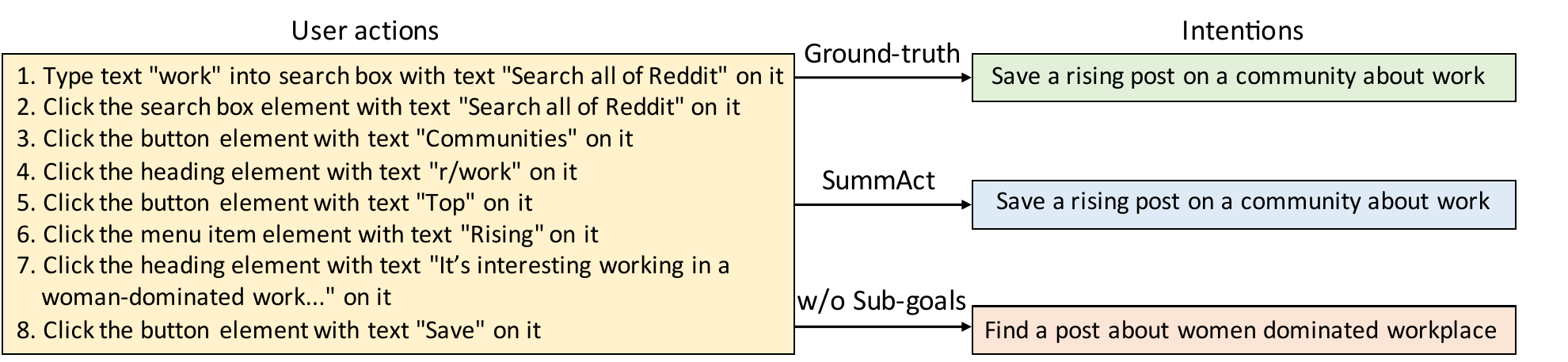}
    \end{subfigure}
    \begin{subfigure}[t]{\linewidth}
        \includegraphics[width=\linewidth]{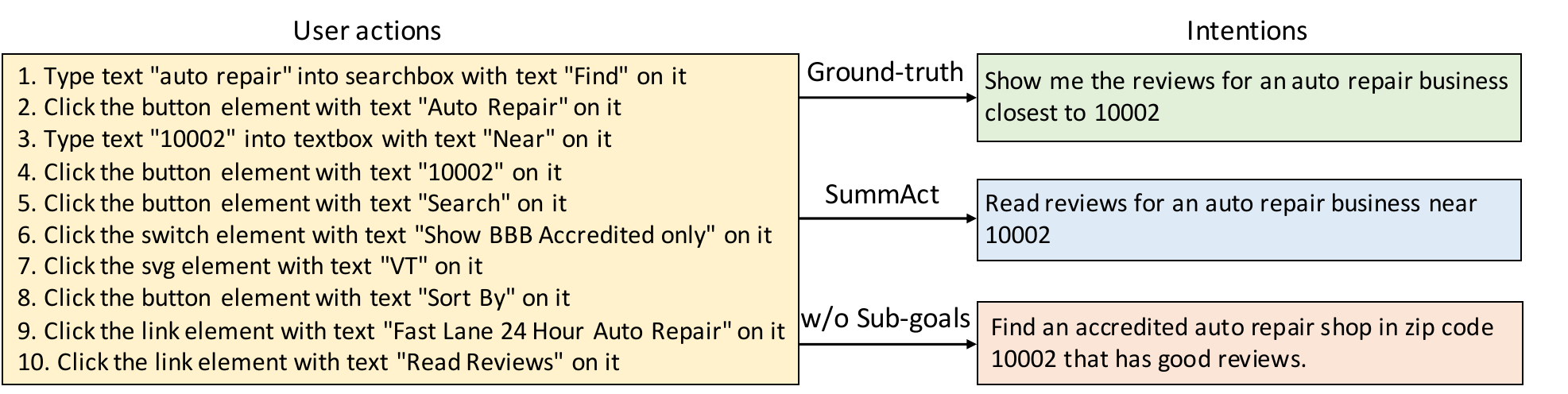}
    \end{subfigure}
    \caption{Two examples showing the input user actions, their underlying ground-truth intentions and those summarised by the full version of \methodName and its ablation removing sub-goals.}
    \Description{X}
    \label{fig:woSubgoal}
\end{figure*}

We then examined the summaries generated by the other ablation (LLM+FT+Attn) in which the sub-goals were removed from \methodName.
We found that the summaries retained specific information but often failed to capture the overarching logic or coherence, especially when handling complex, multi-step interactive behaviour.
\autoref{fig:woSubgoal} shows two examples of this phenomenon, each with the user's input actions and their underlying intentions, the ground-truth intention, the summary generated by the full version of \methodName, and the summary generated by the ablation. %
For example, at the top, the user browsed through top-trending content within a community about work, then picked one post with the heading ``...woman-dominated work...'' and saved it. 
Based on this interactive behaviour, \methodName's summarisation was the same as the ground truth, i.e. \textit{save a rising post on a community about work}.
However, the ablation produced \textit{find a post about women-dominated workplace}, focusing disproportionately on specific content cues, such as the particular post's heading, rather than the overall context of these actions.
This demonstrates that without sub-goals, summaries may lack the essence and broader context of user interactions and instead focus on specific keywords or aspects, resulting in a summary that does not reflect the overall intention.

In the second example, the user searched for \textit{auto repair} with filtering conditions, selected one particular item, and then read its reviews.
This was summarised correctly by \methodName: the summary includes that the user first looks for a business and then verifies its quality according to its reviews.
However, the ablation summarised the intention as \textit{find an accredited auto repair shop in zip code 10002 that has good reviews}, mistakenly understanding the goal of filtering the auto repair business based on their reviews.
This occurs because, without sub-goals, the model processes all behaviour information indiscriminately and struggles to dissect the intricate dependencies and hierarchy among input actions.
As a result, the ablated model erroneously swapped the sequence priorities between \textit{finding auto repair} and \textit{reading reviews}.

This analysis underscores the importance of our hierarchical approach in handling complex interactive behaviour.
By generating intermediate sub-goals, \methodName not only distils key information from different stages of the user actions but also maintains a coherent understanding throughout each interaction stage, ensuring that the final summary encapsulates the overall context.

\section{Applications of Interactive Behaviour Summarisation}

\subsection{Interactive Behaviour Forecasting}
\label{sec:app-NA}

\begin{table*}[ht]
\centering
    \begin{tabular}{lcccccc}
    \toprule
    \multirow{2}{*}{\textbf{Input}} & \multicolumn{2}{c}{\textbf{Cross Domain}} & \multicolumn{2}{c}{\textbf{Cross Task}} & \multicolumn{2}{c}{\textbf{Cross Website}} \\ \cline{2-7} 
     & Element & Operation & Element & Operation & Element & Operation \\ \midrule
     History & 31.2 & 42.1 & 34.2 & 46.2 & 30.6 & 40.4 \\
    History+Summary (\textit{w/o \detailName}) & 37.8 & 45.9 & 43.8 & 47.7 & 34.9 & 43.9 \\
    History+Summary (\textit{Full}) & \textbf{46.8} & \textbf{50.5} & \textbf{47.8} & \textbf{54.5} & \textbf{40.1} & \textbf{45.0} \\ \bottomrule
    \end{tabular}
    \vspace{.3cm}
    \caption{Element accuracy and operation F1 score of next action prediction achieved using 1) only behaviour history, 2) behaviour history plus its summary generated by \methodName \textit{w/o \detailName}, and 3) behaviour history and its summary generated by \methodName full version.
    Both metrics are in percentage.
    The best results are shown in bold.}
    \label{tab:NA-results}
\end{table*}

Interactive behaviour forecasting %
is core to anticipatory and proactive interactive systems~\cite{zhang2022predicting}.
Specifically, we conducted the next action prediction following~\cite{zhang2024mouse2vec}.
Most current next action prediction methods are based solely on the historical information without explicitly understanding overall intentions~\cite{koldijk2012real, kwok2018every, zhang2024dismouse}.
The summarisation of action history can potentially enhance the next action prediction by providing contextual information on users' goal trajectories.

We approached next action prediction as a multi-choice question-answering task, i.e., the model selects from a list of candidate UI elements that users may interact with, following~\cite{chung2024scaling, gu2022don}. %
The pipeline includes three steps~\cite{deng2023mind2web,zheng2024gpt}:
First, we summarise the actions a user has performed using our \methodName.
Then, we employ a candidate extraction method proposed by~\cite{deng2023mind2web} to filter and rank UI elements on the current web page and retain the top-$k$ elements as the candidate targets for the next action.
Retaining only top-$k$ elements is because the raw HTML contains a large amount of noisy UI data %
that can distract LLMs and cause hallucinations~\cite{martino2023knowledge} and exceeds the maximum length of allowed input tokens.
In our experiments, we set $k$ to 50~\cite{deng2023mind2web}.
Finally, we fine-tune an LLM to select the next target UI element out of the 50 candidates and predict its corresponding operation out of three classes (click, select or type).
We used the same fine-tuning set-up as the summarisation, i.e., Mistral-7B as the backbone LLM and the same learning rate, optimiser and scheduler.
We fine-tuned the model for only three epochs, given its fast convergence.
We required the behaviour history to include at least five past actions to offer adequate context, consistent with prior next action prediction works~\cite{kwok2018every,zhang2024mouse2vec}.
The prompt for fine-tuning the LLM included these past actions, the intention summarised by \methodName, and the list of candidate UI elements (see Appendix \ref{sec:prompt-NA}).

We compared our results with two baselines, as shown in ~\autoref{tab:NA-results}.
The first is when only using the action history %
to compare with and examine the effectiveness of summaries in next action prediction.
The other is when using the history plus the summary generated by the ablated version of \methodName excluding the \detailName, to check the specific contribution of keeping UI content as discussed in Section~\ref{sec:method-loss}.
Following~\cite{deng2023mind2web, zheng2023synapse}, we calculated the accuracy of UI element prediction, and F1 score of user operation prediction to measure the imbalanced operation classes.
We showcased the performance of next action prediction on \datasetpc given that this dataset has more variety of intentions and interfaces than \datasetmobile, as shown in Section~\ref{sec:eval-datasets}.

As presented in ~\autoref{tab:NA-results}, integrating summarised intentions consistently enhanced the performance across domains, websites and tasks, with an average 12.9\% %
higher element accuracy and 7.1\% higher operation F1 score.
Moreover, we observed adding the \detailName improved the element accuracy and the operation F1 score by 6.1\% and 4.2\% on average%
, respectively, verifying that the UI content preserved by our method was helpful for next action prediction.

Enhancing next action prediction offers practical benefits for various interactive scenarios.
For instance, the adaptive user interfaces can dynamically adjust the layout and functionality or directly recommend future actions to users to reduce required cognitive and physical demands and enhance usability~\cite{li2024omniactions}.
Additionally, this approach allows automation agents to operate more efficiently by intuitively responding to user preferences without requiring explicit task instructions. %
This potentially leads to smoother, more personalised interactions adapting to evolving user behaviour.

\subsection{Automatic Identification of Behaviour Synonyms}
\label{sec:app-synonyms}

Besides directly capturing user intentions, our summarisation method can also identify ``interactive behaviour synonyms'', which, in our examples, are alternative action sequences reflecting the same underlying user intention.

Practically, we used multiple windows of various lengths from two to the maximum sequence length to respectively segment each input action sequence into sub-sequences.
Our \methodName processed each sub-sequence to summarise its underlying intention.
We then combined all the sub-sequences and calculated the cosine similarity between their sentence embeddings.
Two action sub-sequences are considered as synonyms if their cosine similarity is higher than a threshold.

We showed three example types of synonyms that can provide interesting insights into interaction strategies and system usability:
1) when two synonym behaviours are different but generated from the same UI, the synonyms reflect different strategies or preferences users can take towards an intention; 
2) when two synonym behaviours are different and generated from different UIs, the shorter action path indicates better usability;
3) when two synonym behaviours are generated from different UIs but have same actions, this presents that there are common behaviour patterns or UI designs.

\subsubsection{Different behaviours from the same UI}

\begin{figure*}
    \centering
    \includegraphics[width=\linewidth]{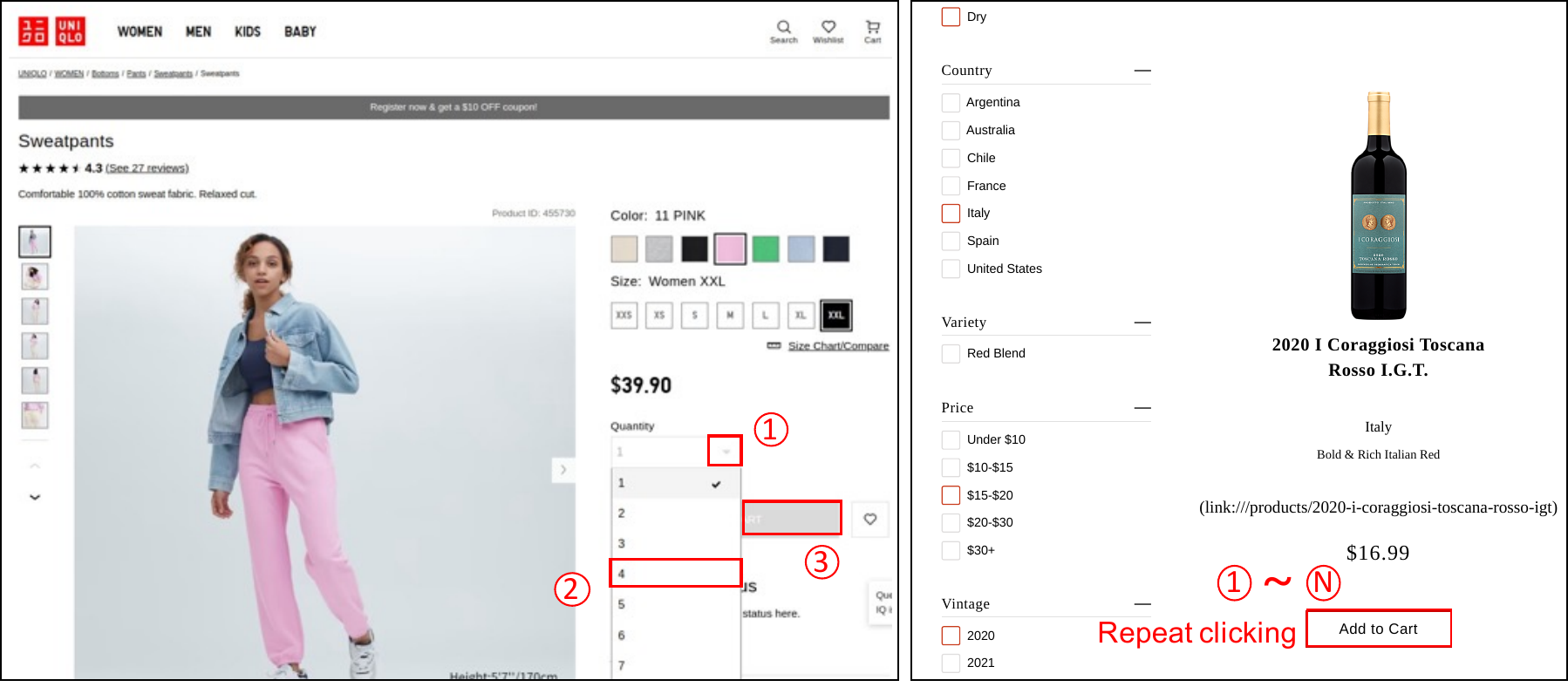}
    \caption{An example of using synonyms to compare UI usability for the task of \textit{adding $N$ items into the shopping cart}.
    The Uniqlo website (left) allows users to add multiple items with just three clicks, while the Macy's website (right) requires one click per item, leading to more effort and less usability as $N$ increases.}
    \Description{X}
    \label{fig:synonym-quant}
\end{figure*}

Based on the same intention on the same UI, users can still generate various behaviours, showing different user preferences, interaction habits and strategies.

For instance, to \textit{add an item to a new shopping list}, users could choose a shorter action path, i.e., creating a new list when adding an item, or a longer trajectory first navigating to the page showing all the existing lists, adding a list there, then returning to the item page, and finally adding the item.
In another example, when the intention was to \textit{find a top-rated restaurant in Miami}, one behaviour directly navigated to a page listing all restaurants and then selected the desired city.
In contrast, another user path first identified the city, browsed a broad range of "things to do", and then narrowed down to restaurants.
These variations may reflect different user preferences, browsing habits, or the clearness of user intention: the former path shows that the user may have a straightforward motivation to look for a \textit{restaurant}. 
At the same time, the latter shows that the user may just look for a \textit{place to go} in the city, not necessarily for a restaurant.
These synonyms can help designers understand user preferences and habits, find the optimal interaction strategies, and design user-tailored interfaces.

\subsubsection{Different behaviours from different UIs}

\begin{figure*}
    \centering
    \begin{subfigure}[t]{\linewidth}
        \includegraphics[width=\linewidth]{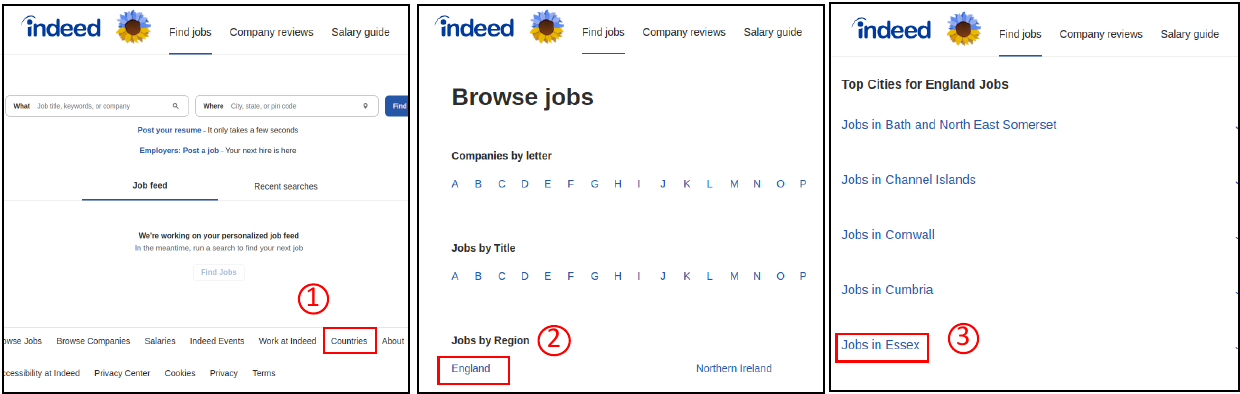}
        \caption{Searching for jobs in Essex on Indeed}
    \end{subfigure}
    \begin{subfigure}[t]{\linewidth}
        \includegraphics[width=\linewidth]{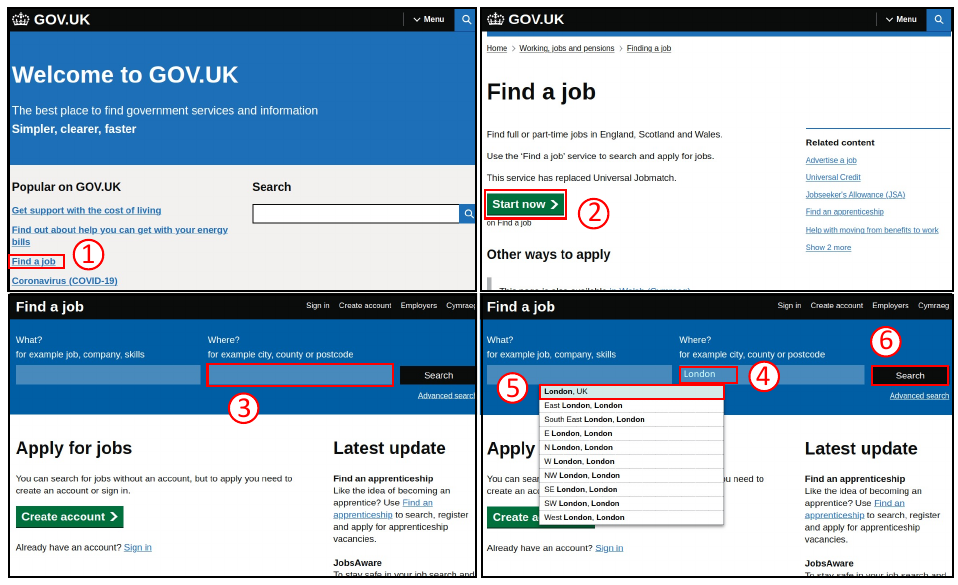}
        \caption{Searching for jobs in London on Gov.UK}
    \end{subfigure}
    \caption{An example of using synonyms to compare UI usability for the intention of \textit{Searching for jobs in city A}.
    On the upper interface, the users can finish the task with only three actions; on the lower interface, the users must perform six actions, indicating worse usability.
    }
    \Description{X}
    \label{fig:synonym-job}
\end{figure*}

UI designs impact the efficiency of achieving interactive intentions, shown by user behaviours.
Therefore, the length of the synonyms found through our summarisation can be used as a metric of UI efficiency.
Unlike classical metrics like the keystroke-level model (KLM) that measure interactive system's usability via task completion time~\cite{frokjaer2000measuring,hornbaek2006current}, this metric will compare UIs via the number of actions required for the same intention.
As shown in ~\autoref{fig:synonym-quant}, to \textit{add $N$ items into the shopping cart}: the Uniqlo website (left) enables users to directly choose the total quantity and add all of them to the cart at once, i.e., finishing with three clicks;
while Macy's (right) only allows adding one item each time, i.e., with $N$ clicks.
When $N$ has a large value, users on the latter interface will have to perform many more actions, harming efficiency and usability.
Another example is shown in~\autoref{fig:synonym-job}, where the intention was to \textit{search for jobs in a city} (Essex in the upper and London in the lower example).
In the upper example, users only needed three clicks to navigate from the main page to view all jobs from the city, i.e., \textit{Countries \textrightarrow England \textrightarrow Jobs in Essex}.
On the contrary, in the lower example, users had to perform more actions to see the list of jobs, i.e., \textit{click on Find a job \textrightarrow click on Start now \textrightarrow click on the text box under Where \textrightarrow type London \textrightarrow click on London \textrightarrow click on Search}.
The reason is that the former website, Indeed, is specifically designed for job searching, optimising its UI to streamline this function.
In contrast, the latter website, Gov.UK, serves multiple functions, not focusing primarily on job hunting, consequently leading to relatively lower efficiency.
As such, these types of synonyms can help UX designers optimise interface design to facilitate quicker and more intuitive access, such as creating shortcuts for the interactive intentions that are dominant among users.

\subsubsection{Same behaviours from different UIs}
We found cases where the synonym behaviours followed the same patterns, although in different interactive systems.
For example, when the intention was \textit{to book a flight ticket from A to B}, user behaviours on different interfaces, including Kayak, Trip.com, American Airlines, and Expedia, typically followed a uniform process: users clicked flights, typed and selected the departure city or airport, and then typed and selected the destination.
When \textit{making an appointment with a doctor} on various medical platforms such as Zocdoc, Mayoclinic.org and Healthgrades, users also employed the same procedure: 
typed and clicked on the type of specialist, browsed and selected an available doctor, and finally picked the appointment time.

Such common patterns in interactive behaviour are why our model can generalise across different websites and tasks (as shown in ~\autoref{tab:eval-results}).
Understanding these patterns also gives UX designers intuitive starting points to create a new interface that aligns with established user behaviour and expectations.
This can reduce the learning curve while ensuring a consistent, friendly user experience.

\subsection{Language-based action retrieval}
\label{sec:app-retrieval}
Through summarisation, \methodName lowers the barrier of understanding and interpreting interactive behaviour. 
Instead of analysing the low-level actions and complex UI contexts, users can directly read their semantic meaning in natural language and further utilise language queries to access specific behaviour trajectories. 
This capability opens up various practical application scenarios.
In \textbf{technical troubleshooting},  the system streamlines the diagnostic process by providing quick access to relevant troubleshooting steps based on user-provided problem descriptions and their intended functions.
In \textbf{educational contexts}, when novice users struggle to complete an interactive goal, the system can quickly pull up the most relevant teaching actions or tutorials from experts for them to watch and learn.
For instance, when users are learning to use complex softwares that require professional skills, such as editing an image with Adobe Photoshop or drawing with AutoCAD, they can type a wanted function or effect, then retrieve, learn and apply other users' action paths implementing it.
Moreover, such retrieval lays the foundation for the development of \textbf{question-answering agents} and \textbf{conversational user interfaces}.
These systems observe and summarise a user behaviour, then retrieve similar behaviours or query related features, such as behaviour frequency from databases %
to provide more context-aware and personalised responses.
For example, based on a task, if we can retrieve a large amount of behaviours, this task may be a common or routine task so that the system can offer automating this task;
furthermore, if the most frequent behaviour has a long action trajectory, the system can adjust the menu and dashboard to reduce users' time and effort.

\section{Discussion}

\subsection{Interactive Behaviour Summarisation}
Although recent works have begun to model interactive behaviour from a natural language perspective~\cite{zhang2023exploring, wen2023empowering, deng2023mind2web}%
, how to use this approach to understand underlying user intentions during interactions remains under-explored.
Our work fills this gap by formulating intention recognition as an interactive behaviour summarisation task, i.e., summarising input action sequences into sentences reflecting latent user intentions.
This formulation represents a paradigm shift in intention modelling.
Traditionally, intention recognition tasks have been constrained by a closed set of predefined intentions, limiting the system's ability to adapt to the varied and evolving needs of users dynamically.
By employing natural language to encapsulate user actions, we offer a more flexible and scalable framework that can intuitively interpret intentions, eliminate the exhaustive definition of every possible user intention, and cater to an open-ended array of these intentions.
The open-ended nature of these summaries allows for recognising unseen, out-of-distribution user intentions, verified by the \methodName's robust performance in the cross-task setting on \datasetpc (see Section ~\ref{sec:eval-quantitative}).
The open-ended set also enables continuous learning and system refinement~\cite{wang2024visionllm}%
, which can integrate emergent user behaviours and preferences without requiring extensive manual updates or reconfigurations.
Furthermore, leveraging a \textit{natural language} representation can improve explainability by grounding the understanding of complex actions within a familiar linguistic framework, lowering the barrier of interpreting and analysing user behaviour.

In addition to its inherent utility, we also demonstrate various example applications enabled by the summary, including interactive behaviour forecasting,
automatic behaviour synonym identification,
and %
language-based behaviour retrieval.
Understanding users' intentions from past actions is important to forecast future actions along the intention.
Therefore, we enhanced current computational next action prediction methods based on action history by adding the summary as an additional input.
\methodName's summarisation led to better prediction performance, %
as shown in ~\autoref{tab:eval-results}.
These improvements underscore the effectiveness of adding semantic information through interactive behaviour summarisation.
Moreover, if two behaviours have the similar summary%
, they can be considered synonyms, i.e., alternatives achieving the same goal.
Synonyms can provide insights into understanding diverse user preferences and interaction strategies, optimising interactive systems, and finding common patterns in behaviour and UI designs across systems.
The linguistic nature of the generated summaries also unlocks language-based action retrieval, meaning that users can query and retrieve specific action sequences using simple language requests.
This will facilitate learning from expert users in navigating complex interfaces or completing difficult tasks and enable question-answering agents and conversational user interfaces.

\subsection{Design of \methodName}
We propose a novel LLM-based approach, \methodName, that includes two methodological contributions and design choices to generate natural language summarisation from user actions and their UI context.
The first design choice is to employ a hierarchical process adept at handling the complexity and variability of user actions.
The method first enriches the semantic context available for the final summary generation by producing intermediate sub-goals from low-level actions. 
The second design choice is an \detailName mechanism applied on the fine-tuning loss to prevent over-abstraction by preserving essential context information embedded in the UI elements, such as the meaning of the clicked button.
Extensive quantitative comparisons with ablations showed that both design choices contribute to the effectiveness of \methodName (see ~\autoref{tab:eval-results}).
Additionally, in Section~\ref{sec:eval-qualitative}, the qualitative evaluations examined the generated text and verified that both designs were necessary and complementary to each other: without the sub-goals, the summaries can have logical errors; while without the detail attention, the summaries lacked action details (see \autoref{fig:woSubgoal}). %

Moreover, as shown in ~\autoref{tab:eval-results}, our method has significantly improved upon the results obtained from using pretrained %
LLMs (up to three times better, on \datasetmobile), which is for now a common practice in HCI research that use LLMs~\cite{huang2024automatic, liu2024unblind, wang2023enabling}.
The substantial enhancement brought by \methodName reveals the considerable potential of further enhancements in these HCI works and positions our approach as a pioneering example in the field.
Our findings suggest that the HCI community can gain immensely by further adapting advanced natural language processing techniques to specific HCI applications.

In implementing \methodName, we utilised Mistral-7B as the backbone LLM due to its lightweight and robust performance across various NLP tasks.
However, our \methodName framework allows replacing Mistral-7B with other LLM models if computing resources are available or more powerful models appear.%

\subsection{Limitations and Future Work}
In our work, we investigated summarising the interactive behaviour at the UI-element level, i.e., the users operate on UI elements such as buttons or combo boxes.
In the future, we will dive deeper into the raw, pixel-level interactive behaviours, e.g., integrating the specific on-screen locations of each move, click or tap into our model, which will carry more information about users and their intentions~\cite{zhang2024mouse2vec,zhang2024dismouse}.
Currently, we integrated UI information via HTML and DOM elements.
Future enhancements can adopt vision language models to process GUI screenshots~\cite{zheng2023synapse,li2023blip}.
This will allow for capturing visual cues that HTML alone cannot provide, such as iconography, layout spatial arrangements, and thematic designs that influence user interactions.
The proposed \detailName implemented a crude match between tokens by simply considering whether they are identical, but moving forward, we could consider matching their semantics to increase the model's flexibility.
Moreover, our exploration of interactive behaviour summarisation is based on the assumption that the observed UI trajectories accurately reflect user intentions without error.
While essential for developing our method, in real-world interactions, it is possible that actions may not always convey true intentions due to errors in user operation, which will be interesting for developing more robust models in the future.

\section{Conclusion}
In this work, we model interactive behaviour from a natural language viewpoint and investigate a novel interactive behaviour summarisation task, summarising input actions into natural language descriptions.
These descriptions reflect user intentions underlying their interactive behaviour.
Towards this task, we propose \methodName~-- an LLM-based method with two specific designs, hierarchical summarisation and \detailName.
We evaluated our method on two datasets, covering both the web and mobile interactive settings, from both the quantitative and qualitative perspectives.
Results demonstrated the effectiveness of our method in summarising intentions and the complementary contributions of our two designs.
We then showcased example applications of interactive behaviour summarisation, including behaviour forecasting, automatic identification of behaviour synonyms, and language-based behaviour retrieval.
The natural language representation of interactive behaviour can boost the explainability of computational behaviour modelling and contribute to developing more intuitive and responsive interactive systems.
Furthermore, our significant improvement over the common practice of directly using LLMs in HCI suggests large potential benefits from further adapting advanced NLP techniques to HCI tasks.

\bibliographystyle{ACM-Reference-Format}
\bibliography{main}

\appendix
\section{Describing \datasetpc Actions in Natural Language}
\label{sec:appendix-description}

As mentioned in Section ~\ref{sec:eval-datasets}, we preprocessed the original actions provided by \datasetpc into natural language descriptions to better leverage the understanding and reasoning capabilities inherent in LLMs.
\autoref{tab:action-example} presents three examples of the original actions, each from an user operation category (click, select or type).
\begin{table*}[htb]
    \begin{tabular}{ccccc}
    \hline
    \textbf{\begin{tabular}[c]{@{}c@{}}Element \\ Category\end{tabular}} & \textbf{\begin{tabular}[c]{@{}c@{}}Element \\ Content\end{tabular}} &  & \textbf{\begin{tabular}[c]{@{}c@{}}User \\ Operation\end{tabular}} & \textbf{\begin{tabular}[c]{@{}c@{}}Element \\ Content (Additional)\end{tabular}} \\ \hline
    $[button]$ & Add to Cart & $\to$ & CLICK & -- \\
    $[combobox]$ & Sort By & $\to$ & SELECT & Price Low to High \\
    $[searchbox]$ & Search & $\to$ & TYPE & Johannesburg \\ \hline
    \end{tabular}
    \vspace{.2cm}
    \caption{Three example input action strings from \datasetpc dataset.
    Every string, representing an input action, contains the information of the interacted UI element (category and content) and user's operation on it.} %
    \label{tab:action-example}
\end{table*}

Following ~\cite{niu2024screenagent}, we first split the original action strings to the UI element category, content (the inherent meaning of this element, e.g., its name or the text on it), additional content for type and select operations (specific content the user is interested in, e.g., selected value from a combo box) and the user operation category; and then inserted these components into the natural language template, structured as:
\begin{small}
\begin{align*}
    &\text{If}\; Operation=CLICK: \; [Operation] \; \text{the}\; [Category]\; \text{element with text ``} [Content] \text{'' on it}\\
    &\text{If}\; Operation=SELECT: \; [Operation] \; \text{``}[Content(Additional)]\text{'' from} \; [Category]\; \text{with text ``} [Content] \text{'' on it}\\
    &\text{If}\; Operation=TYPE: \; [Operation] \; \text{text}\; \text{``}[Content(Additional))]\text{''}\; \text{into}\; [Category] \text{with text ``} [Content] \text{'' on it}\\
\end{align*}
\end{small}
Through this templating approach, the example action strings are represented as: 
\begin{itemize}[leftmargin=15pt]
    \item Click the button element with text ``Add to Cart'' on it
    \item Select ``Price Low to High'' from combobox with text ``Sort By'' on it
    \item Type text ``Johannesburg'' into searchbox with text ``Search'' on it
\end{itemize}

\section{Prompts}

\subsection{In-Context Learning Prompts for Sub-goal Generation}
\label{sec:prompt-subintention}
\begin{figure*}
    \centering
    \includegraphics[width=\linewidth]{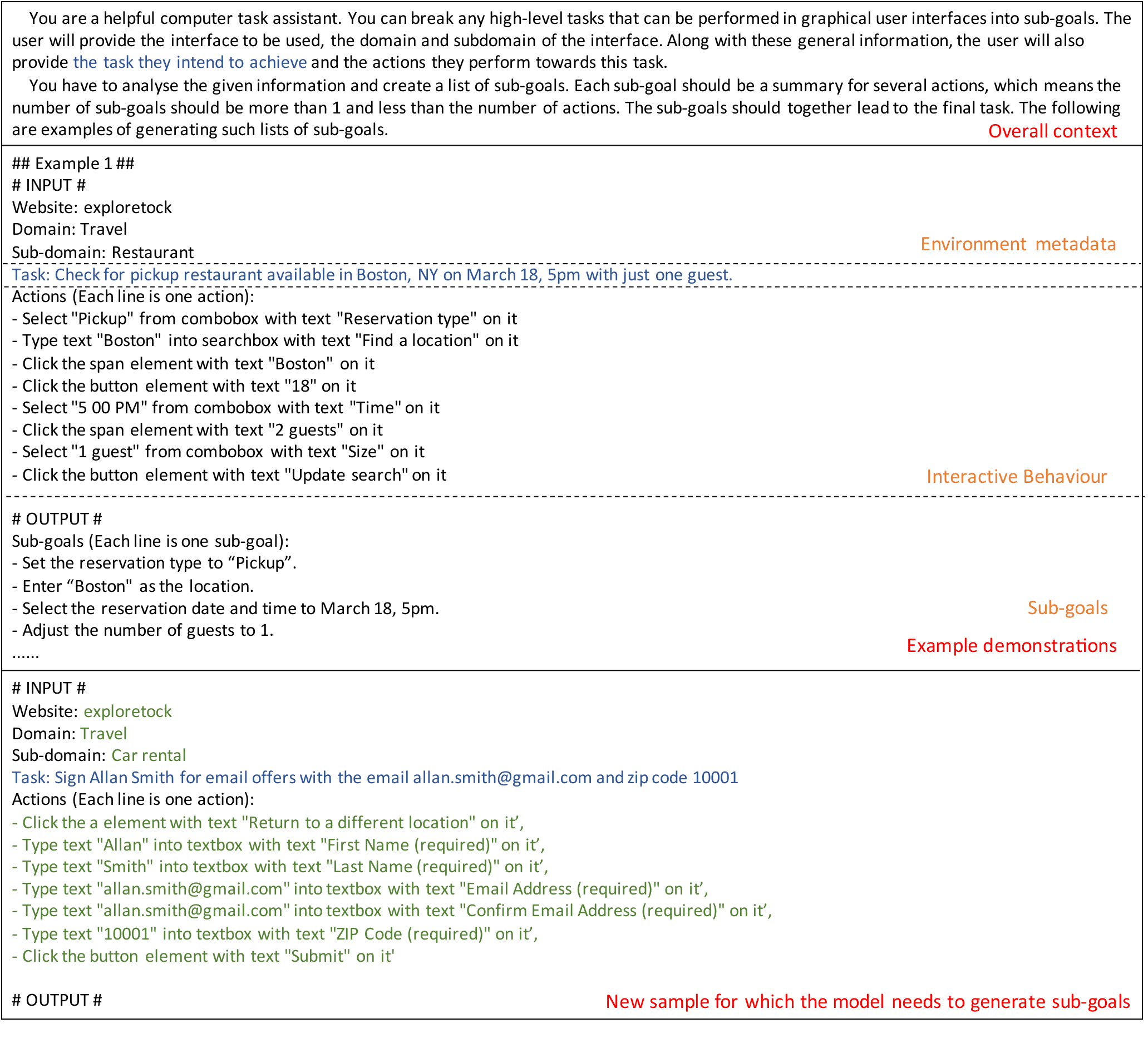}
    \caption{Prompt used to generate sub-goals using in-context learning.}
    \Description{X}
    \label{fig:prompt-subintention}
\end{figure*}
As described in Section~\ref{sec:method-subintention}, we apply in-context learning to generate sub-goals from low-level interactive actions.
\autoref{fig:prompt-subintention} shows an example of the prompt we used for the pretrained LLM.
The prompt has three main components:
1) an overall context describing the task the LLM needs to solve, the input format and expected output;
2) examples used in the in-context learning annotated by experts, each including environment metadata, task (only in training samples), interactive behaviour and sub-goals (we only show one example due to its excessive length);
3) and a new sample for which the model needs to generate sub-goals, where the text in green should be replaced and updated for each sample.
The texts in blue are related to the ground-truth overall intentions, and thus should be removed from testing samples to avoid data leakage.
The example in ~\autoref{fig:prompt-subintention} is from \datasetpc dataset, which provides the metadata including website, domain and sub-domain. 
When using the prompt on \datasetmobile dataset, simply replace them with the provided mobile application as the new meta data.

\subsection{Prompts for Detail Enhanced Fine-Tuning}
\label{sec:prompt-finetuning}
\begin{figure*}
    \centering
    \includegraphics[width=.8\linewidth]{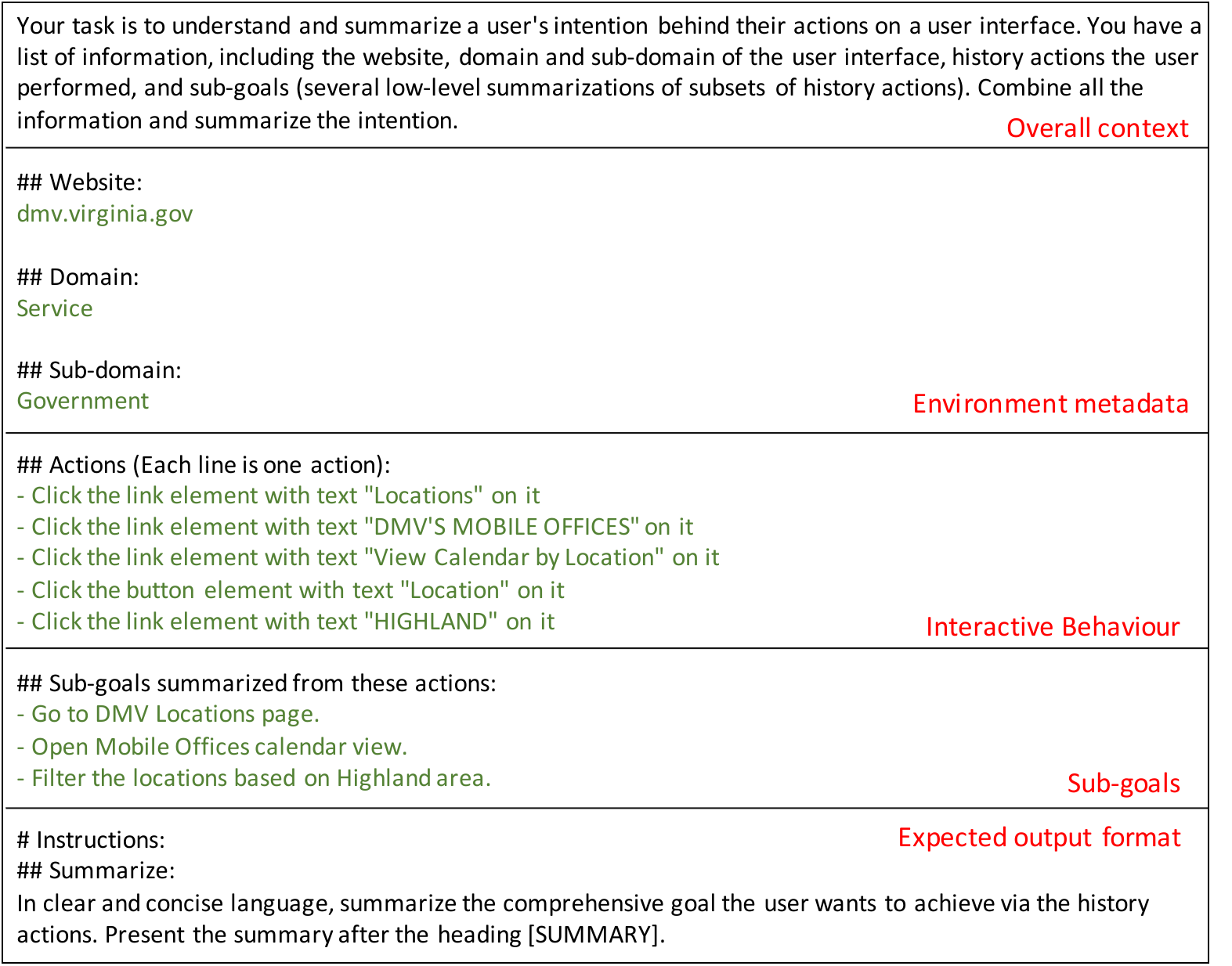}
    \caption{Prompt used to summarise the overall intention from low-level interactive behaviour and mid-level sub-goals.}
    \Description{X}
    \label{fig:prompt-finetune}
\end{figure*}
As described in Section~\ref{sec:method-loss}, we fine-tune LLMs to summarise the final, overall intention using both low-level interactive behaviour and mid-level sub-goals.
\autoref{fig:prompt-finetune} illustrates an example of the prompt we used for fine-tuning.
The prompt comprises five parts:
1) an overall context describing the task the LLM needs to solve, the input format and expected output;
2) environment metadata;
3) user input actions;
4) sub-goals;
and 5) expected output format.
The text in green should be replaced and updated for each sample.
The example in ~\autoref{fig:prompt-finetune} is from \datasetpc dataset, which provides the metadata including website, domain and sub-domain. 
When using the prompt on \datasetmobile dataset, simply replace them with the provided mobile application as the metadata.

\subsection{Prompts for Next Action Prediction}
\label{sec:prompt-NA}
\begin{figure*}
    \centering
    \includegraphics[width=\linewidth]{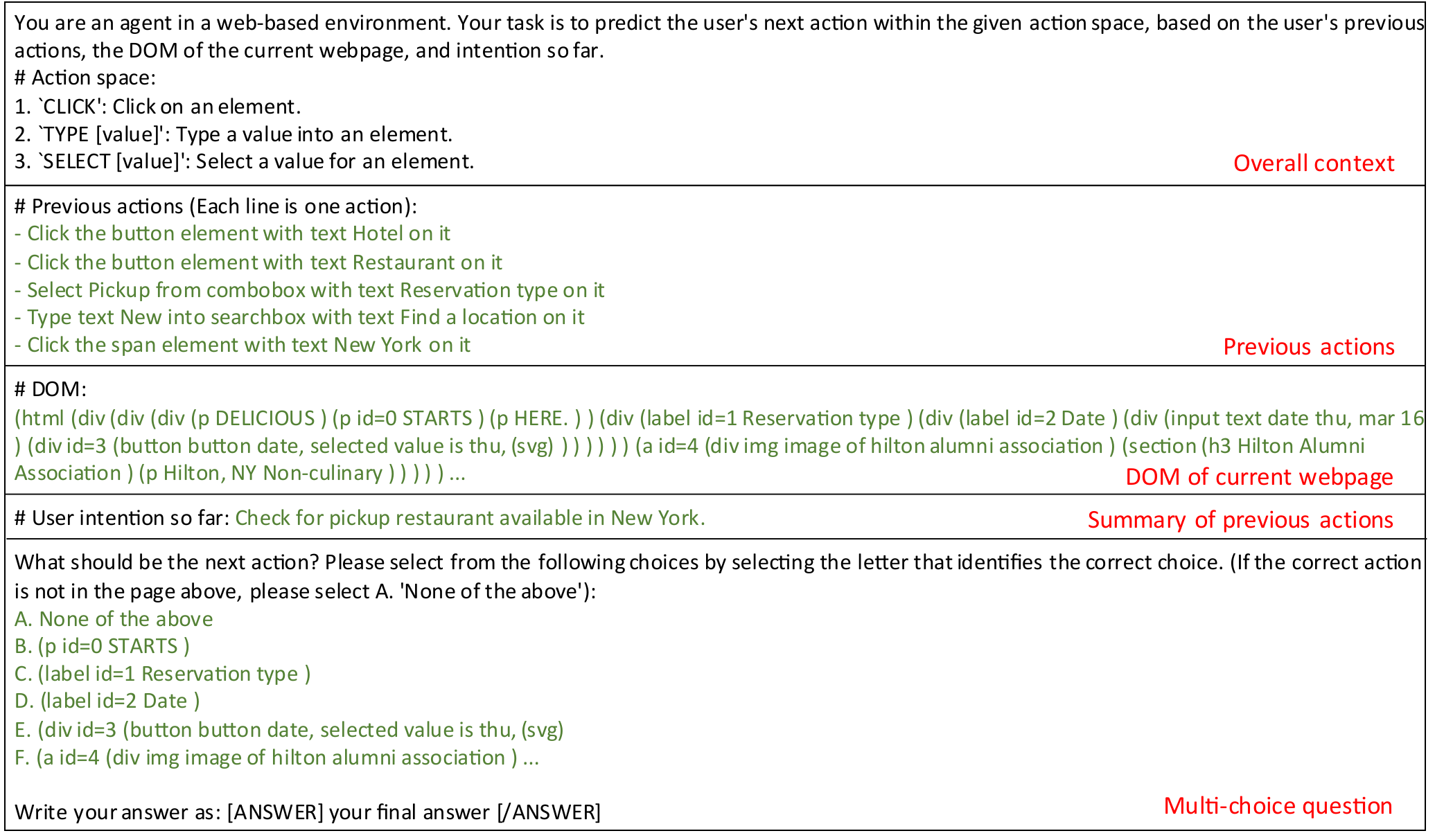}
    \caption{Prompt used to forecast interactive behaviour based on previous input actions and the current web page.
    }
    \Description{X}
    \label{fig:prompt-NA}
\end{figure*}

As described in Section~\ref{sec:app-NA}, we fine-tune an LLM to predict the next action a user may perform based on the history actions and the intentions summarised from them via our \methodName. 
\autoref{fig:prompt-NA} shows an example of this prompt, consisting of five parts:
1) an overall context describing the task the LLM needs to solve and the allowed action space;
2) input actions the user has performed;
3) DOM of the current webpage related to candidate UI elements;
4) intention summarised from the performed actions;
and 5) candidate UI elements as the next target.
The text in green should be replaced and updated for each sample.

\end{document}